\documentclass[sigconf,authorversion]{acmart}
\usepackage{float}
\usepackage{tabularx}
\usepackage{bm}
\usepackage{enumitem}
\AtBeginDocument{%
  }
\copyrightyear{2026}
\acmYear{2026}
\setcopyright{cc}
\setcctype{by}
\acmConference[L@S '26]{Proceedings of the Thirteenth ACM Conference on Learning @ Scale}{June 29-July 03, 2026}{Seoul, Republic of Korea}
\acmBooktitle{Proceedings of the Thirteenth ACM Conference on Learning @ Scale (L@S '26), June 29-July 03, 2026, Seoul, Republic of Korea}
\acmDOI{10.1145/3774398.3811580}
\acmISBN{979-8-4007-2293-6/2026/06}

\begin{document}
\title[Push and Pull in Community College Cross-Enrollment]{Push and Pull in Community College Cross-Enrollment: Remoteness, Articulation, and Student Mobility}
\author{Conrad Borchers}
\orcid{0000-0003-3437-8979}
\affiliation{
    \institution{Carnegie Mellon University}
    \city{Pittsburgh}
    \state{PA}
    \country{USA}
}
\email{cborcher@cs.cmu.edu}

\author{Robin Schmucker}
\orcid{0000-0002-5275-3608}
\affiliation{
    \institution{University of California, Berkeley}
    \city{Berkeley}
    \state{CA}
    \country{USA}
}
\email{schmucker@berkeley.edu}

\author{Ashutosh Tiwari}
\orcid{0000-0002-3720-0853}
\affiliation{
    \institution{University of California, Berkeley}
    \city{Berkeley}
    \state{CA}
    \country{USA}
}
\email{ashutosh.tiwari@berkeley.edu}

\author{Zachary A. Pardos}
\orcid{0000-0002-6016-7051}
\affiliation{
    \institution{University of California, Berkeley}
    \city{Berkeley}
    \state{CA}
    \country{USA}
}
\email{pardos@berkeley.edu}
\renewcommand{\shortauthors}{Conrad Borchers, Robin Schmucker, Ashutosh Tiwari, and Zachary A. Pardos}
\newcommand{\revisioncrc}[1]{\textcolor{black}{#1}}

\begin{abstract}
Cross-enrollment across institutions can expand access to courses and support student progression. Still, little is known about how geographic constraints and institutional policies jointly shape cross-enrollment within community college (CC) systems. We adopt a push--pull framework: geographic remoteness constrains feasible cross-institution mobility, while credit mobility may attract enrollment expressed as \emph{articulation} (CC-to-university: credit toward a four-year partner) and \emph{course equivalencies} (CC-to-CC: equivalencies across the system). Using de-identified administrative records from a 12-institution community college system (100,547 students; 1,290,311 course enrollments), we quantify outgoing and incoming cross-enrollment and relate these patterns to institutional remoteness and credit mobility. We find that less remote colleges exhibit higher outgoing and incoming cross-enrollment than more remote colleges. Further, cross-enrolled students are more likely to take articulated courses, and institutions with higher equivalency ratios receive higher incoming cross-enrollment (8.62\% vs. 6.70\%). This association was slightly stronger at more remote colleges. This study demonstrates how analysis of complex college systems can surface factors shaping student mobility and inform the design of cross-enrollment and articulation policies in CC systems.
\end{abstract}

\maketitle

\section{Introduction}

Community college (CC) students navigate complex pathways as they pursue credentials, manage constrained course availability, and prepare for transfer to four-year institutions \cite{morales2022policy,gonzalez2025community}. \emph{Cross-enrollment} allows students to enroll in courses at partner institutions while remaining matriculated at their home college. It can expand course availability and scheduling flexibility, potentially reducing delays and barriers to graduation \cite{sutton2018guide,grubb2017answer}.

Participation in cross-enrollment, however, is uneven across institutions and students. Prior research documents common barriers, including limited student awareness, administrative complexity, and uncertainty around credit transfer \cite{morales2022policy}. This evidence is largely qualitative and interview-based, with relatively small samples, so it cannot characterize system-level patterns. Existing quantitative work has focused on predicting enrollment decisions from student characteristics, behavior, and academic interests \cite{wang2017analyzing,chaturapruek2021studying}. Such approaches, in turn, do not capture the structural and contextual factors emphasized in qualitative work (e.g., geographic constraints and institutional arrangements). Empirical evidence linking these structural conditions to cross-enrollment at scale remains limited.

These challenges raise a central question: under what conditions does cross-enrollment support student progress and academic mobility, and for whom? Geographic context is one underexamined dimension. It directly shapes the feasibility of in-person cross-enrollment \cite{acton2025distance}. Prior work suggests that access alone is insufficient to explain student outcomes. Although online courses can expand access, community college students tend to benefit most from a balanced mix of modalities, and predominantly online enrollment is often associated with lower completion rates \cite{shea2018online}. In-person enrollment thus remains a key component of effective course-taking pathways, motivating closer examination of how geographic accessibility constrains or enables cross-enrollment within CC systems.

Similar push--pull frameworks have been applied to how students choose their university, including in international mobility contexts \cite{mazzarol2002push}. We adopt this lens to examine cross-enrollment behavior within CC systems. From a push perspective, geographic remoteness may constrain feasible cross-enrollment opportunities by increasing travel time and limiting access to nearby partner institutions. Colleges located farther from peers may therefore exhibit lower rates of both outgoing and incoming cross-enrollment. From a pull perspective, two distinct factors may attract cross-enrollment: \emph{articulation} (CC-to-university agreements that grant credit toward a four-year partner) and \emph{course equivalencies} (CC-to-CC recognition of course equivalence across the system). Both can increase the perceived value of cross-enrolled coursework by reducing uncertainty about credit applicability \cite{morales2022policy,hill2025understanding,laviolet2020beyond}. These pull factors may be especially salient for geographically remote institutions, where fewer nearby options make the transfer payoff of a cross-enrollment decision more attractive to students.

We draw on de-identified administrative records from a statewide CC system to examine how geographic remoteness and articulation agreements jointly shape cross-enrollment patterns. We contribute a push--pull framework for horizontal mobility, system-level quantitative evidence on how remoteness and both pull factors (articulation and course equivalencies) relate to cross-enrollment flows, and implications for the design of scalable transfer pathways. We ask:

\noindent \textbf{RQ1:} Push: How does geographic remoteness relate to outgoing and incoming cross-enrollment patterns across CCs?

\noindent \textbf{RQ2:} Pull: Are CCs with stronger articulation (CC-to-university) or stronger course equivalency offerings (CC-to-CC) more likely to attract incoming cross-enrolled students?

\section{Related Work}

Administrative data create new opportunities for Learning at Scale research to move beyond single-course or single-institution analyses and examine how entire systems structure learner pathways. Prior work has leveraged large-scale data to model transfer, predict course-taking, and improve articulation through computational methods \cite{fischer2020mining,pardos2019data}. Most studies of student mobility, however, focus on \emph{vertical transfer} from CCs to four-year institutions \cite{anderson2006effectiveness}. Far less attention has been paid to \emph{horizontal mobility} within multi-institution CC systems, where students cross institutional boundaries without formally transferring. Such patterns (sometimes described as swirling or double-dipping \cite{mccormick2003swirling}) represent system-level learning behavior that unfolds across colleges rather than within them. Geographic access has been linked to transfer in rural contexts \cite{higgins1999relationship}, and articulation agreements are well established as predictors of completion \cite{roksa2008credits}; quantitative evidence on how these structural factors shape cross-enrollment flows at scale is lacking. This work integrates spatial analytics with articulation and course equivalency data across an entire statewide system, reframing cross-enrollment as a system-level learning phenomenon and advancing a broader Learning at Scale agenda by understanding how institutional design, geography, and policy interact to shape learner mobility, access to coursework, and the structure of educational pathways.

\section{Data}
\label{sec:data}

This study uses de-identified administrative student- and course-level records from a statewide CC system in the U.S. consisting of 12 CC institutions. The dataset spans multiple academic years and captures student enrollment histories, cross-enrollment activity across institutions, articulation (CC-to-university) with a single partner four-year state university, and course equivalency (CC-to-CC) data across the system. All data were de-identified before analysis in accordance with an approved IRB protocol.

The full dataset contains records for 100,547 students and 2,019 distinct courses, totaling 1,290,311 course enrollments. Among all students, 12,979 (12.9\%) engaged in at least one instance of cross-enrollment, defined as enrolling in a course at a CC other than their home institution. At the enrollment level, 103,505 enrollments (8.0\%) correspond to cross-enrollment activity. Students completed an average of 10.09 semesters (median = 9.0, IQR = 6.0).

\subsection{Cross-Enrollment Definitions}

Cross-enrollment here denotes \emph{horizontal mobility} within the CC system: students enroll in courses at institutions other than their primary (home) CC while remaining enrolled in the system. A student is cross-enrolled if they take at least one course at a non-home CC during their enrollment history. We examine cross-enrollment from both sending and receiving perspectives. Outgoing cross-enrollment indicates whether students from a given home institution enroll elsewhere in the system; incoming cross-enrollment indicates whether an institution receives enrollments from students whose home institution is another CC. 

\subsection{Articulation and Course Equivalencies}

We examine two distinct pull factors. \emph{Articulation} refers to CC-to-university recognition: whether a CC course is formally recognized for credit toward degree requirements at a partner four-year university. Of the 2,019 courses offered across the system, 1,456 (72.1\%) have an active articulation agreement. At the course level, each enrollment is flagged as articulated or non-articulated. \emph{Course equivalencies} refer to CC-to-CC recognition of course equivalence across the system (distinct from articulation). We compute an \emph{equivalency ratio} for each CC as the proportion of its distinct courses that have course equivalencies; this measure is used to examine whether colleges with stronger equivalency offerings attract higher levels of incoming cross-enrollment. 

\subsection{Geographic Measures and Remoteness}

To capture geographic accessibility, we compute pairwise driving distances between all CC pairs using the Google Distance Matrix API. For each CC, we derive a \emph{Remoteness Index} as the average driving distance to its three closest institutions, approximating each college’s local opportunity set for feasible commuting-based cross-enrollment while avoiding sensitivity to a single nearest neighbor. Institutions are categorized into high- and low-remoteness groups via a median split on the Remoteness Index (given its skewed distribution), yielding a coarse but interpretable measure of geographic constraint for system-level analysis.

\subsection{Unit of Analysis}

Analyses are conducted at the course-enrollment and institutional levels. For RQ1, we analyze institutional-level outgoing and incoming cross-enrollment rates by geographic remoteness. For RQ2, we examine both pull factors: articulation at the course-enrollment level (articulated vs. non-articulated enrollments) and course equivalencies at the institutional level (equivalency ratio and incoming cross-enrollment rates). This structure supports examination of spatial push factors and both articulation and equivalency as pull factors. \revisioncrc{Binary splits were performed because of the limited sample size of institutions in our sample}.

\section{Analytical Methods}

We use the constructs defined in Section~\ref{sec:data} (cross-enrollment rates by sending and receiving institutions, the Remoteness Index, course-level articulation, and the institutional equivalency ratio). The analyses below are designed to answer each research question directly.

\subsection{RQ1: Remoteness and Cross-Enrollment}

RQ1 asks how geographic remoteness relates to outgoing and incoming cross-enrollment. To answer, we compare cross-enrollment rates between low- and high-remoteness institution groups. For \emph{outgoing} cross-enrollment, we aggregate to the student level (whether each student had at least one cross-enrollment) and then by home institution; for \emph{incoming}, we use enrollment-level counts by institution. For each outcome, we build a $2 \times 2$ contingency table (remoteness group $\times$ cross-enrollment yes/no) and test for independence using Pearson's $\chi^2$. We report proportions with 95\% Wilson score intervals and odds ratios (OR) with 95\% CIs to quantify the association between remoteness and cross-enrollment.

\subsection{RQ2: Articulation and Cross-Enrollment}

RQ2 asks whether colleges with stronger articulation or stronger course equivalency offerings attract more incoming cross-enrolled students. We address it using both pull factors. \emph{Course-level (articulation, CC-to-university):} We test whether courses recognized for credit at a partner four-year institution are overrepresented among cross-enrolled enrollments by comparing the odds that an enrollment is articulated for cross-enrolled vs. non-cross-enrolled enrollments. We stratify by remoteness and report the ratio of odds ratios to assess whether this association differs by remoteness context. \emph{Institution-level (course equivalencies, CC-to-CC):} We split colleges by \emph{mean} equivalency ratio and compare incoming cross-enrollment rates between high- and low-equivalency groups ($\chi^2$, OR). We also replicate the analysis using a median split for robustness. Finally, we report the Pearson correlation between equivalency ratio and incoming cross-enrollment rate across the 12 institutions. 

\section{Results}

\begin{figure}[t]
    \centering
    \includegraphics[width=\linewidth]{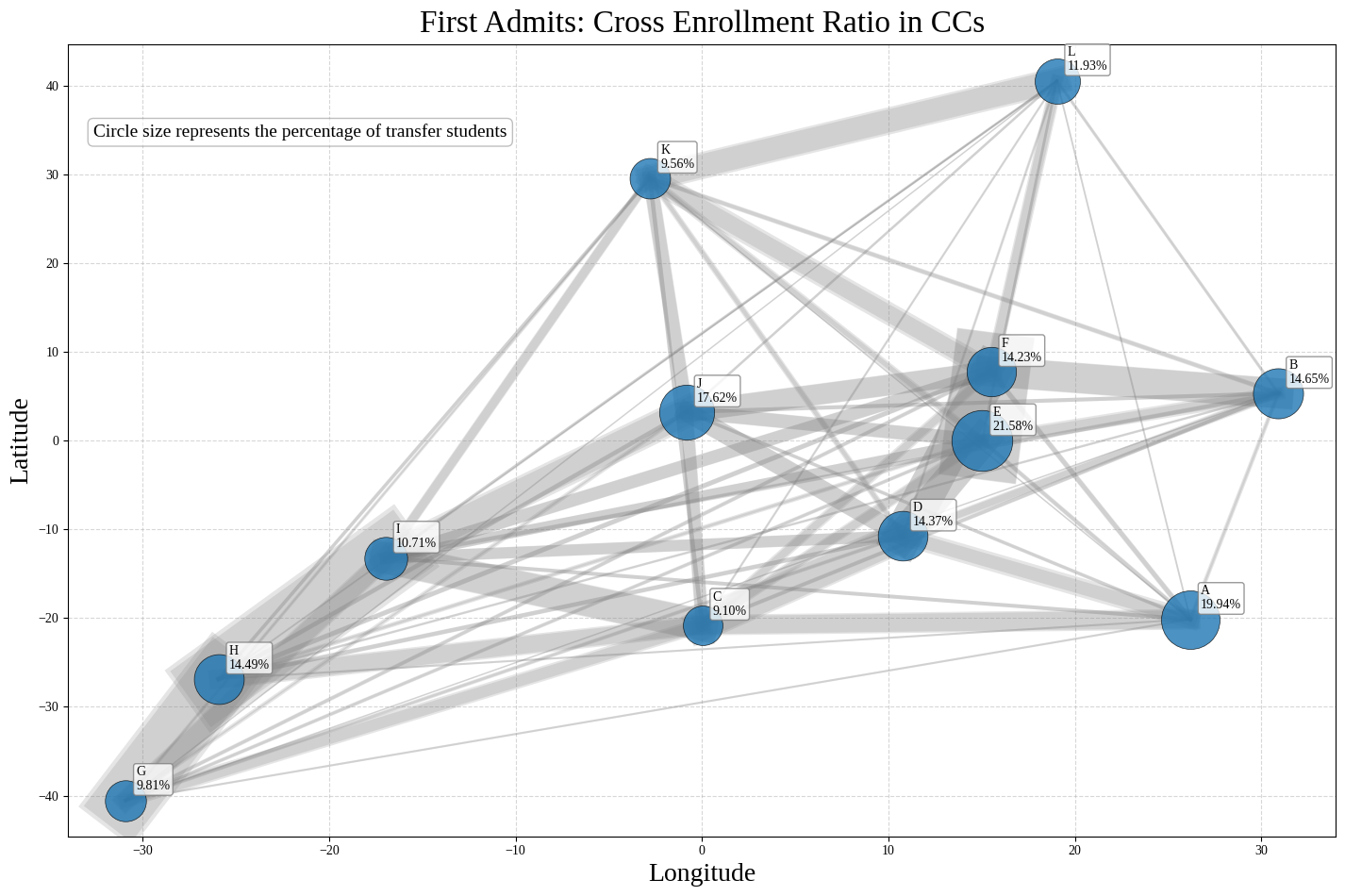}
    \caption{Geographical plot of outgoing and incoming cross-enrollment rates across CCs\revisioncrc{ represented by circles, with lines indicating individual cross-enrollment decisions}.}
    \label{fig:cross_enrollment_map}
\end{figure}

Figure~\ref{fig:cross_enrollment_map} summarizes outgoing and incoming cross-enrollment patterns across the CC system. Substantial variation appears across institutions, with higher rates concentrated among geographically proximate colleges. 

\subsection{RQ1: Geographic Remoteness and Cross-Enrollment Patterns}

We test whether institutional remoteness is associated with outgoing and incoming cross-enrollment.

\paragraph{Outgoing Cross-Enrollment}
Institutions classified as less remote exhibit substantially higher rates of outgoing cross-enrollment than those classified as more remote. Students enrolled at less remote colleges were more likely to take at least one course at another CC (Low-remoteness: 14.50\%, 95\% CI [14.22, 14.79]; High-remoteness: 10.72\%, 95\% CI [10.43, 11.02]). A $\chi^2$ test of independence confirms that this difference is statistically significant, $\chi^2(1) = 312.31$, $p < .001$. The corresponding odds ratio indicates that students at less remote institutions have 1.41 times higher odds of engaging in outgoing cross-enrollment than students at more remote institutions (OR = 1.41, 95\% CI [1.36, 1.47]).

\paragraph{Incoming Cross-Enrollment}
Incoming cross-enrollment follows a similar pattern. Less remote institutions receive a higher proportion of cross-enrolled students (Low-remoteness: 9.38\%, 95\% CI [9.31, 9.45]; High-remoteness: 6.12\%, 95\% CI [6.06, 6.19]). The difference is statistically significant, $\chi^2(1) = 4512.26$, $p < .001$, with OR = 1.59 (95\% CI [1.57, 1.61]), indicating substantially higher odds of incoming cross-enrollment at less remote colleges.

Taken together, these results indicate that geographic remoteness is strongly associated with both sending and receiving cross-enrollment behavior, consistent with a push mechanism in which physical accessibility shapes feasible cross-institution mobility.

\subsection{RQ2: Articulation and Course Equivalencies as Pull Factors}

We test whether articulation (CC-to-university) and course equivalencies (CC-to-CC) are associated with incoming cross-enrollment. Both are analyzed as pull factors.

\paragraph{Articulation (CC-to-University)}
At the course level, cross-enrolled students are more likely than non-cross-enrolled students to enroll in articulated courses (i.e., courses recognized for credit at a partner four-year institution). The association is statistically significant ($p < .001$), with OR = 1.14, indicating approximately 14\% higher odds that a cross-enrolled enrollment involves an articulated course. Articulated coursework is thus disproportionately represented among cross-enrollment activity.

When stratified by institutional remoteness, this relationship remains consistent. At more remote institutions, the odds of cross-enrolled students taking articulated courses are slightly higher (OR = 1.16) than at less remote institutions (OR = 1.12). The ratio of odds ratios (ROR) of 1.03 indicates a modestly stronger association between articulation and cross-enrollment in more remote settings, though the magnitude of this difference is small.

\paragraph{Course Equivalencies (CC-to-CC)}
At the institutional level, we use the \emph{equivalency ratio} (proportion of each college's distinct courses with course equivalencies; this is not articulation). We split colleges by the \emph{mean} equivalency ratio into high- and low-equivalency groups. Colleges with higher equivalency ratios receive higher levels of incoming cross-enrollment. Institutions above the mean equivalency ratio exhibit a higher incoming cross-enrollment rate (8.62\%, 95\% CI [8.56, 8.68]) than those below the mean (6.70\%, 95\% CI [6.62, 6.78]). This difference is statistically significant, $\chi^2(1) = 1383.71$, $p < .001$. 
Using a median split on the equivalency ratio yields a smaller, though robust, effect (OR = 0.99, 95\% CI [0.97, 1.00], $p = .033$).
Institutions with weaker equivalency offerings have lower odds of incoming cross-enrollment (OR = 0.76, 95\% CI [0.75, 0.77], $p < .001$). Correlation across the 12 institutions between equivalency ratio and incoming cross-enrollment rate is moderate and positive (Pearson $r = 0.54$) but not statistically significant ($p = .069$), likely due to the small number of institutions.

\section{Discussion}

This study addresses a gap in the cross-enrollment literature by providing system-level, quantitative evidence on how geographic and institutional factors jointly shape cross-enrollment in CC systems. Prior work has documented barriers through qualitative accounts \cite{morales2022policy} or focused on individual-level prediction \cite{wang2017analyzing}, leaving limited empirical understanding of how spatial and policy factors structure enrollment. We integrate driving-distance-based remoteness with articulation and course equivalency data and apply a push--pull framework to administrative records across a statewide system.

Addressing \textbf{RQ1}, geographic remoteness is associated with both outgoing and incoming cross-enrollment: less remote colleges exhibit higher rates of students enrolling elsewhere and receive more incoming cross-enrolled enrollments, indicating that physical accessibility shapes feasible cross-institution mobility. For \textbf{RQ2}, we find two pull effects: at the course level, articulation (CC-to-university) is overrepresented among cross-enrolled enrollments; at the institution level, colleges with higher equivalency ratios (course equivalencies, CC-to-CC) attract more incoming cross-enrolled students. Spatial constraints thus limit opportunity sets, while both pull factors---articulation and course equivalencies---appear to influence institutional attractiveness within those constraints. Practically, system and institutional leaders can use remoteness to identify where cross-enrollment is likely constrained by geography and use articulation and equivalency data to target outreach or pathway design where pull effects may be strongest (e.g., at more remote colleges, where the articulation--cross-enrollment association is slightly stronger).

Learning at Scale emphasizes systems that serve many learners across institutions and the use of data to understand pathways \cite{pardos2019data,chaturapruek2021studying}. The theoretical contribution is a push--pull framework for horizontal mobility that is operationalizable from institutional records (e.g., articulation and course equivalency data). The empirical contribution is, to our knowledge, the first system-level, quantitative evidence that geographic remoteness and both pull factors (articulation and course equivalencies) jointly shape cross-enrollment flows. Prior L@S and related work have addressed course-to-course articulation \cite{pardos2019data} and individual course consideration at scale \cite{chaturapruek2021studying}, but not cross-enrollment as a system-level phenomenon with both spatial and policy dimensions. We do not claim that push and pull are the only forces or that our design supports causal inference. Instead, we show that they are associated with observed flows and that the approach aligns with a Learning at Scale agenda focused on system-level analytics and transfer pathways.

Regarding limitations, findings are based on a single 12-institution system. Generalizability to other states or configurations is unknown. The small number of institutions limits power for institution-level correlations. Future work should examine heterogeneity by student characteristics (e.g., major, transfer intent, and background), link cross-enrollment to transfer and completion outcomes, and test whether articulation, course equivalencies, or advising interventions amplify pull effects. \revisioncrc{Similarly, future work may explore how course characteristics not captured in our data, such as online course offerings, tie into student cross-enrollment decisions.}

\section{Conclusion}

Using a push--pull framework, we examined how geographic accessibility and two distinct pull factors---articulation (CC-to-university) and course equivalencies (CC-to-CC)---shape horizontal mobility within a multi-institution community college system. We provide system-level quantitative evidence that remoteness (push) and both pull factors structure cross-enrollment flows. Colleges located closer to other colleges both send and receive more cross-enrolled students, indicating that geographic remoteness constrains feasible mobility. Articulation functions as a pull factor: cross-enrolled students disproportionately take articulated courses, and this association is slightly stronger in more remote settings. Course equivalencies also function as a pull factor: institutions with higher equivalency ratios (mean split) attract more incoming cross-enrollment. When geographic options are limited, both articulation and equivalency appear to become more salient.

In practice, articulation agreements and course equivalencies are administratively costly to maintain, but cross-enrollment can increase enrollment by leveraging existing course capacity. While remoteness creates structural disadvantage, expanding articulation and equivalency offerings may partially offset it by increasing institutional attractiveness. For Learning at Scale research, this study reframes cross-enrollment as a system-level, networked phenomenon measurable from administrative data. Future work should develop predictive models of cross-enrollment flows and examine links to transfer and completion outcomes to inform resource optimization across institutions.

\begin{acks}
This research was supported by the Bill and Melinda Gates Foundation (Award \#91974). The views expressed in this research are those of the authors and do not necessarily reflect the positions of the Gates Foundation.
\end{acks}

\bibliographystyle{ACM-Reference-Format}
\bibliography{main}

\end{document}